\documentclass[english,showpacs,twocolumn,superscriptaddress]{revtex4-1}
\usepackage[T1]{fontenc}
\usepackage[latin9]{inputenc}
\usepackage{amsmath}
\usepackage{amssymb}
\usepackage{graphicx}

\makeatletter
\usepackage[colorlinks,
            linkcolor=blue,
            anchorcolor=blue,
            citecolor=blue
            ]{hyperref}

\makeatother

\usepackage{babel}
\begin{document}
\title{Measure the temperature of the spin via weak measurement }
\author{Yusuf Turek}
\email{yusufu1984@hotmail.com}

\affiliation{School of Physics and Electronic Engineering, Xinjiang Normal University,
Urumqi, Xinjiang 830054, China}
\begin{abstract}
In this study, we give a new proposal to measure the temperature of
the spin in a sample in magnetic resonance force microscopy system
by using postselected weak measurement, and investigate the fisher
information to estimate the precision of our scheme. We showed that
in high temperature regime the temperature of the spin can be measured
via weak measurement technique with proper postselection and our scheme
able to increase the precision of temperature estimation.
\end{abstract}
\pacs{03.65.Ta, 42.50.-p,06.20.Dk,05.70.-a}
\maketitle

\section{Introduction}

Measurement is a basic concept in physics and any information of the
system can be obtained from the measurement. Unlike the classical
measurement, in most of the quantum measurement processes the information
of the system only can be obtained or estimated by using indirect
measurements. Using quantum measurement terminology, in a quantum
measurement process, there have an interaction between measuring device
(probe) and measured system, and for guarantee the measurement precision
the interaction time must be very short so that the system and the
probe itself doesn't effect the results of measurements. In mathematics,
these requirements of quantum measurement can be expressed by von
Newmann Hamiltonian $H=g\hat{A}\otimes\hat{P}$\cite{Neumann}. Here,
$\hat{A}$ is the operator of measured system we want to measure,
$\hat{P}$ the canonical momentum of measuring device, and $g$ represent
the measuring strength. If the interaction strength between the system
and the probe is strong, i.e. $g\gg1$, we can get the information
of the system by single trial with very small error, the Stern-Gerlah
experiment is a typical example. In contrast, if the interaction strength
between probe and system is very small, i.e. $g\ll1$, there have
still interference between different eigenvalues of the system observable
we want to measure, and after single trial we can not obtain the information
of the system precisely. 

However, the weak measurement, as a generalized von Neumann quantum
measurement theory, was proposed by Aharonov, Albert, and Vaidman
in 1988 opens a new route to measure the system information in weak
coupling measurement problems\cite{Aharonov(1988)}. In weak measurement
technique, the coupling between the pointer and the measured systems
is sufficiently weak,and the obtained information by single trial
is trivial, but using its inherent feature one can get enough information
as precisely as possible\cite{Tollaksen(2010)}. One of the distinguished
properties of weak measurement compared with strong measurement is
that its induced weak value of the observable on the measured system
can be beyond the usual range of the eigenvalues of that observable\cite{qparadox(2005)}.
The feature of weak value usually referred to as an amplification
effect for weak signals rather than a conventional quantum measurement
and used to amplify many weak but useful information in physical systems\cite{Dressel(2014),Nori,Shikano(2010)}.
So far, the weak measurement technique has been applied in different
fields to investigate very tiny effects, such as beam deflection\cite{Pfeifer(2011),Hosten2008,Hogan(2011),Zhou(2013),Starling(2009),Dixon(2009)},
frequency shifts\cite{Starling(2010)-1}, phase shifts\cite{Starling(2010)},
angular shifts\cite{Magana(2013),Bertulio(2014)}, velocity shifts\cite{viza(2013)},
and even temperature shift\cite{Egan(2012)}. Furthermore, it has
been applied to solve some fundamentals of quantum physics such as
quantum paradoxes(Hardy's paradox\cite{Aharonov(2002),Lundeen(2009),Yokota(2009)}
and the three-box paradox\cite{Resch(2004)}), quantum correlation
and quantum dynamics\cite{Aharonov(2005),Aharonov(2008),Holger(2013),Shikano(2012),Aharonov(2011),Shikano(2011)},
quantum state tomography\cite{Lundeen(2012),Braveman(2013),Kocsis(2011),Malik(2014),Salvail(2013),Lundeen(2011)},
violation of the generalized Leggett-Garg inequalities\cite{Palacios,Suzuki(2012),Emary(2014),Goggin(2011),Groen(2013),Dressel(2011)}
and the violation of the initial Heisenberg measurement-disturbance
relationship\cite{Lee(2012),Eda(2014)},etc. 

We know that temperature is a basic concept in thermodynamics and
most of the properties of a matter is directly related with its temperature.
According to the thermodynamics the temperature is an intensive quantity
and independent to sample's volume and mass. Thus, it is possible
to measure the temperature of nanoscale objects which inserted in
large sample. The motion of nanoscale objects obey the quantum mechanics
and with the recent progress to manipulation of individual quantum
systems, it has been possible to temperature readings with nanometric
spatial resolution\cite{Neu,Kubo,Toyli}. Quantum thermometery also
applied to precisely estimate the temperature of fermonic\cite{Hauck,Haupt}
and bosonic\cite{White,Braun} hot reservoirs, micromechanical resonators\cite{Bruelli2,Brunelli,Higgins}
and nuclear spins\cite{Raitz}. The nanoscale system is too fragile
and measuring the temperature of such systems requires that the measurement
process should not disturb them to much while keeping the high measurement
precision. Furthermore, as investigated in previous studies\cite{Dressel(2014)},
the weak measurement almost doesn't destroy the measured system and
can get the desired system information via statistical techniques
with high precision. Thus,here raising an intriguing question as to
whether one can measure the temperature of nanoscale objects by using
postselected weak measurement technique. In recent study\cite{Pati(2019)},
they investigated the method to measure the temperature of a bath
using the postselected weak measurement scheme with a finite dimensional
probe, and anticipated more applications of post weak measurement
method in thermometry. 

In this paper, motivated by the previous study\cite{Pati(2019)},
we investigate to measure the temperature of nanoscale system(spin)
which inserted in large thermal reservoir by using postselected weak
measurement. We assume that there have a weak interaction between
spin and cantilever in magnetic resonance force microscopy(MRFM) system\cite{Berman}.
We show that the temperature of the spin which stayed in hot reservoir
with termpreature $T$, can be measured by postselected weak measurement
technique with proper postselected states of the spin, and the weak
value can be read out from optical experimental processes. In temperature
estimation process the fluctuation always exists due to the indirect
measurement, but the quantum estimation theory provides the tool to
evaluate lower bounds to the amount of fluctuations for a given measurement.
To investigate the precision of our method, we evaluate the fisher
information(FI) for the estimation of temperature via postselected
weak measurement. We found that in high temperature regime the FI
is more larger than unity with proper postselected states of the spin,
and this enable us to show that postselected meak measurement method
is indeed cane be used in temperature estimation process in MRFM system
with high presicion. 

The rest of the paper is organized as follows. In Sec. II, we give
the setup for our system. In Sec. III, we give the details how to
measure the temperature of a spin via postselected weak measurement.
In Sec. IV, we investigate the Fisher information to investigate the
precision of our scheme.We give a conclusion to our paper in Sec.
V. 

\section{Model setup}

\begin{figure}
\includegraphics[width=8cm]{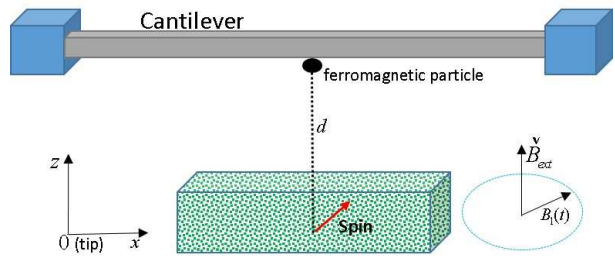}

\caption{\label{fig:fig1}(Color online) The setup of our scheme. A spin and
the cantilever are coupled with each other by a tiny ferromagnetic
particle attached to the middle of the cantilever. $d$ is the distance
between the bottom of the ferromagnetic particle and the spin at equilibrium
position, $z=0$. }
\end{figure}

As illustrated in Fig.\ref{fig:fig1}, two ends of a cantilever used
in MRFM is fixed and a small ferromagnetic particle is attached at
the middle of the cantilever\cite{Berman}. The force produced by
a single spin on the ferromagnetic particle effects the parameters
of the mechanical vibrations of the cantilever. There have a external
permanent magnetic field $B_{ext}$ directed in the $z$ direction
exerts on the spin and a ferromagnetic particle attached on the middle
of the cantilever also exerts a gradient dipole magnetic field produced
by the ferromagnetic particle on the spin $\text{\ensuremath{\partial B_{d}}/\ensuremath{\partial z}}$
with $B_{d}=\frac{2}{3}\mu_{0}M_{0}(\frac{R_{0}}{R_{0}+d+z})^{3},$
where $\mu_{0}$ is the permeability of the free space, $d$ is the
distance between the bottom of the ferromagnetic particle and the
spin is initially at equilibrium position($z=0$), $R_{0}$ and $M_{0}$
are the radius and magnetization of the ferromagnetic particle,respectively.
The spin also exposed to the transversal frequency modulated rf magnetic
field $\text{\ensuremath{\vec{B}_{1}(t)=B_{1}\left(\cos(\omega t+\phi(t)),\sin(\omega t+\phi(t)),0\right)} }$in
$x-y$ plane. The motion of the cantilever with ferromagnetic particle
can be modeled as a simple harmonic oscillator with effective mass
$m$ and frequency $\text{\ensuremath{\omega_{c}}}$. we can express
the oscillation position $z$ of the cantilever with creation and
annihilation operators $\hat{a}$ and $\hat{a}^{\dagger}$ by $\hat{z}=\sigma(\hat{a}+\hat{a}^{\dagger})$,
with $\text{\ensuremath{\sigma=\sqrt{\text{\ensuremath{\hbar}}/2m\omega_{c}}}}$.
In the rotating system coordinates(RSC) the Hamiltonian of this system
can be written as\cite{Berman} 
\begin{align}
H & =\hbar\omega_{c}\hat{a}^{\dagger}\hat{a}+\hbar\omega_{z}\hat{S}_{z}+\hbar\omega_{R}\hat{S}_{x}-\hbar g\hat{S}_{z}\otimes\hat{z}.\label{eq:Hamil}
\end{align}
Here, $\omega_{R}=\gamma B_{1}$, $\omega_{z}=\gamma B_{0}-\omega-\frac{d\phi}{dt}$
, $g=\gamma\frac{\partial B_{d}}{\partial z}$, $\gamma$ is the gyromagnetic
ratio and $B_{0}=B_{ext}+B_{d}^{(0)}$ is the total magnetic field
on the spin when $z=0$ in $z$ direction, and the gradient of the
dipole field $\text{\ensuremath{\partial B_{d}}/\ensuremath{\partial z} }$is
taken at the spin location when $z=0$. $\hat{S}_{z,x}$are defined
by Pauli operators $\hat{S}_{x}\rightarrow\frac{1}{2}\hat{\sigma}_{x}$
and $\hat{S}_{z}\rightarrow\frac{1}{2}\hat{\sigma}_{z}$, respectively.
As we can see, our coupling system Hamiltonian consists of three parts(see
Eq.\ref{eq:Hamil}); first term represents the motion of cantilever,
middle two terms describes the spin system and we denote with $H_{s}$,
and the final term describes the interaction of spin and cantilever,
and we denote it with $H_{int}$. According to the technique of MRFM,
if the frequency of the spin oscillations matches the resonance frequency
of the cantilever vibrations, the spin force will amplify the vibrations
of the cantilever and its motion can be detected by optical methods\cite{Berman}.
In this study, we will focus on to measure the temperature of the
single spin which stayed in reservoir with temperature $T$ in MRFM
system. Hereafter, in this paper we use the unit $\hbar=1$.

\section{Measure the temperature of the spin}

As given in Eq.(\ref{eq:Hamil}), the interaction Hamiltonian between
the system(spin) and pointer(cantilever) is given by the standard
von Neumman Hamiltonian\cite{qparadox(2005)}
\begin{equation}
H_{ini}=-g\hat{S}_{z}\otimes\hat{z}\label{eq:ini}
\end{equation}
 Here, $g$ is a coupling coefficient between spin and cantilever,
and its value depends on gyromagnetic ratio $\text{\ensuremath{\gamma}}$
and the gradient magnetic field at the spin position produced by ferromagnetic
particle. $\hat{z}=\int z\vert z\rangle\langle z\vert dz$ is the
position operator,while the conjugate momentum operator is $\text{\ensuremath{\hat{p}=\int p\vert p\rangle\langle p\vert dp}}$
where $[\hat{x},\hat{p}]=i\text{\ensuremath{\hat{I}}}$. We consider
the spin as system and cantilever as measuring device(pointer) to
measure the temperature of the spin. We assume that initially the
system (spin) is in a heat bath with temperature $T$, and reached
the thermal equilibrium state $\text{\ensuremath{\rho}}_{s}=e^{-\beta H_{s}}/Tr(e^{-\beta H_{s}})$.
Here $\beta=1/k_{B}T$ , $k_{B}$ is Boltzmann constant and it taken
as unity hereafter. We assume that the initial state of measuring
device and system are prepared as $\text{\ensuremath{\phi(z)=\langle z\vert\phi\rangle}}$
and $\text{\ensuremath{\rho(0)=\rho_{s}}}$, respectively. Then, under
the action of unitary evolution operator $U(t)=\exp(-i\int_{0}^{t}H_{ini}d\tau)$,
the initial state of total system $\rho(0)=\rho_{s}\otimes\vert\phi\rangle\langle\phi\vert$
will be evolved to 
\begin{equation}
\text{\ensuremath{\rho(t)=U^{\dagger}(t)\rho(0)U(t)}}.\label{eq:rho}
\end{equation}
This is the state of the total system after interaction, but we only
interested on final state of measuring device(cantilever) which contains
system information. If we assume that the interaction strength $g$
is very small , then it is enough to consider the expansion of unitary
operator up to its first order, and Eq.(\ref{eq:rho}) becomes as
\begin{equation}
\text{\ensuremath{\rho(t)\approx\rho_{s}\otimes\vert\phi\rangle\langle\phi\vert+igt\left[\hat{S}_{z}\otimes\hat{z},\rho_{s}\otimes\vert\phi\rangle\langle\phi\vert\right].}}\label{eq:Tim}
\end{equation}
 According the standard terminology and procedure of weak measurement
theory, if we choose the state $\text{\ensuremath{\vert\psi_{f}\rangle}}$
as the final state of the spin, and project it onto the total system
state, Eq.(\ref{eq:Tim}), it gives the unnormalized final state of
measuring device as
\begin{equation}
\text{\ensuremath{\phi(t)\approx\langle\psi_{f}\vert\rho_{s}\vert\psi_{f}\rangle e^{igt\text{\ensuremath{\hat{S}}}_{w}\hat{z}}\vert\phi\rangle\langle\phi\vert e^{-igt\hat{S}_{w}\hat{z}}.}}\label{eq:finm}
\end{equation}
Here, 
\begin{equation}
\text{\ensuremath{S_{w}}=\ensuremath{\frac{\langle\psi_{f}\vert\hat{S}_{z}\rho_{s}\vert\psi_{f}\rangle}{\langle\psi_{f}\vert\rho_{s}\vert\psi_{f}\rangle}}}\label{eq:WV}
\end{equation}
 is the weak value of spin $z$ component and it related to the initial
state of the spin which reached thermal equilibrium with a heat bath
with temperature $T$. As can see from Eq.(\ref{eq:finm}) and Eq.(\ref{eq:WV}),
we can deduce the spin temperature by properly choosing the final
state$\text{\ensuremath{\vert\psi_{f}\rangle}}$. However, we have
to note that $\text{\ensuremath{\vert\psi_{f}\rangle}}$ cannot be
eigenstate of operator $\hat{S}_{z}$, otherwise the temperature will
not related to the weak value and our scheme will lose its validity
to measure the temperature of the spin. 

To read the temperature of the spin system, next we will study the
property of weak value of spin operator $\hat{S}_{z}$ by assuming
that it in the hot bath with high temperature, i.e. $\beta\ll1$.
Furthermore, we assume that the eigenvalues of $\hat{S}_{x}$ and
$\hat{S}_{y}$ are $\epsilon_{m}$and $\varepsilon_{m}$, and the
corresponding eigenstates are $\vert\psi_{m}\rangle$ and $\vert\varphi_{m}\rangle$,
i.e.,$\hat{S}_{x}\vert\psi_{m}\rangle=\epsilon_{m}\vert\psi_{m}\rangle$
and $\text{\ensuremath{\hat{S}_{z}\vert\varphi_{m}\rangle=\varepsilon_{m}\vert\varphi_{m}\rangle}}$,
respectively. Then, the weak value of the spin $z$ component can
be rewritten as 
\begin{align}
S_{w} & =\frac{\langle\psi_{f}\vert\hat{S}_{z}e^{-\beta H_{s}}\vert\psi_{f}\rangle}{\langle\psi_{f}\vert e^{-i\beta H_{s}}\vert\psi_{f}\rangle}\nonumber \\
 & =\frac{\sum_{n,m}e^{-\beta(\omega_{z}\varepsilon_{n}+\omega_{R}\epsilon_{m})}\langle\psi_{n}\vert\phi_{m}\rangle\langle\psi_{f}\vert\text{\ensuremath{\hat{S}_{z}}}\vert\psi_{n}\rangle\langle\phi_{m}\vert\psi_{f}\rangle}{\sum_{n,m}e^{-\beta(\omega_{z}\varepsilon_{n}+\omega_{R}\epsilon_{m})}\langle\psi_{n}\vert\phi_{m}\rangle\langle\psi_{f}\vert\psi_{n}\rangle\langle\phi_{m}\vert\psi_{f}\rangle}\nonumber \\
 & =\frac{\langle\psi_{f}\vert\hat{S_{z}}\vert\psi_{f}\rangle-\beta\langle\psi_{f}\vert\hat{S_{z}}\hat{H}_{s}\vert\psi_{f}\rangle}{1-\beta\langle\psi_{f}\vert\hat{H}_{s}\vert\psi_{f}\rangle}.\label{eq:WV7}
\end{align}
During the derivation of Eq.(\ref{eq:WV7}), we use $e^{\pm x}\approx1\pm x$,
$(x\ll1)$, and completeness of the basis $\vert\psi_{m}\rangle,\text{\ensuremath{\vert\varphi_{m}\rangle,}i.e. \ensuremath{\text{\ensuremath{\sum_{m}\vert\psi_{m}\rangle\langle\psi_{m}\vert=\sum_{m}\vert\varphi_{m}\rangle\langle\varphi_{m}\vert=1.}}}}$Since
$(1-x)^{-1}\approx1-x,(x\ll1)$, we can rewrite the Eq.(\ref{eq:WV7})
as below 
\begin{equation}
S_{w}=\left[\langle\psi_{f}\vert\hat{S_{z}}\vert\psi_{f}\rangle-\beta\langle\psi_{f}\vert\hat{S_{z}}\hat{H}_{s}\vert\psi_{f}\rangle\right]\left(1-\beta\langle\psi_{f}\vert\hat{H}_{s}\vert\psi_{f}\rangle\right).\label{eq:Wv8}
\end{equation}
 From this relation, we can express the temperature parameter $\beta$
by using the weak value $S_{w}$ as 
\begin{equation}
\text{\ensuremath{\beta=\frac{S_{w}-\langle\hat{S}_{z}\rangle}{Conv(S_{z},H_{s})-\langle\psi_{f}\vert\hat{S}_{z}\hat{H_{s}}\vert\psi_{f}\rangle},}}\label{eq:Beta}
\end{equation}
where $Conv(S_{z},H_{s})=\langle\hat{S}_{z}\rangle\langle\psi_{f}\vert\hat{H}_{s}\vert\psi_{f}\rangle$
and $\langle\hat{S}_{z}\rangle=\langle\psi_{f}\vert\hat{S}_{z}\vert\psi_{f}\rangle$,
respectively. From Eq.(\ref{eq:Beta}) we can see that since the value
of $\langle\hat{S}_{z}\rangle$ and $Conv(S_{z},H_{s})$ can be find
by using postselection $\vert\psi_{f}\rangle$, if we can find the
value of $S_{w}$, then the temperature of the spin can be estimated
very easily. For example, if we take $\vert\psi_{f}\rangle=\frac{1}{\sqrt{2}}(\vert\psi_{1}\rangle+\vert\psi_{2}\rangle)$,
then Eq.(\ref{eq:Beta}) reduced to 
\begin{equation}
\beta=\frac{2(1-2S_{w})}{\omega_{R}+3\omega_{z}}.\label{eq:Beta2}
\end{equation}
From the above theoretical results we can deduce that the estimation
of spin in MRFM system depends on the postselected weak value. Thus,
in the remaining part of this section we will discuss the possibility
to find the weak value of our scheme. We can assume that the motion
of the cantilever can be described by simple harmonic oscillator,
and its initial state can be written as
\begin{equation}
\phi_{0}\left(z\right)=\langle z\vert0\rangle=\left(\frac{1}{2\pi\sigma^{2}}\right)^{\frac{1}{4}}\exp\left(-\frac{z^{2}}{4\sigma^{2}}\right)
\end{equation}
where $\sigma$ is the width of the oscillation beam. The normalized
final state of the cantilever after postselection, i.g. Eq.(\ref{eq:finm}),
can be written as 
\begin{equation}
\vert\phi_{f}\rangle=\text{\ensuremath{\kappa}}[\vert0\rangle+ig_{0}S_{w}\vert1\rangle],\label{eq:Finm}
\end{equation}
where $g_{0}=gt\sigma$, $\text{\ensuremath{\kappa=(1+g_{0}^{2}\vert S_{w}\vert^{2})^{-\frac{1}{2}}}}$
and we the $\phi_{0}(z)=\langle z\vert0\rangle$ and $\phi_{1}(z)=\langle z\vert1\rangle$
represent the ground and first excited states of catilever, respectively.
The expectation values the position and momentum operators under the
final state of the catilever $\text{\ensuremath{\vert\phi_{f}\rangle}}$
can be calculated as 
\begin{align}
\langle z\rangle & =-2\sigma g_{0}\kappa^{2}\Im S_{w}
\end{align}
 and 
\begin{align}
\langle p\rangle & =\frac{\kappa^{2}g_{0}}{\sigma}\Re(S_{w})
\end{align}
 respectively. According to the results of recent studies\cite{Yutaka(2014),Lundeen(2019)},
we can measure the real and imaginary parts of the weak values with
optical experiments. Thus, in our scheme the temperature of spin can
be measured in the Lab with proper optical experimental setups. 

\section{Precision of our scheme}

Fisher information is the maximum amount of information about the
parameter that we can estimate from the system. For a pure quantum
state $\text{\ensuremath{\vert\phi\rangle}}$, the quantum fisher
information estimating the parameter $\beta$ is 

\begin{equation}
F(\beta)=4[\langle\partial_{\beta}\text{\ensuremath{\phi}}\vert\partial_{\beta}\phi\rangle-\vert\langle\phi\vert\partial_{\beta}\phi\rangle\vert^{2}],\label{eq:Fisher}
\end{equation}
where the state $\text{\ensuremath{\vert\phi\rangle}}$is the normalized
final state of the cantilever, i.e. Eq.(\ref{eq:Finm}). The variance
of unknown parameter $\text{\ensuremath{\triangle\beta}}$ is bounded
by the Cramér-Rao bound 
\begin{equation}
Var\left(\beta\right)\ge\frac{1}{NF(\beta)},\label{eq:VarB}
\end{equation}
where $N$ is the number of measurements. From this definition of
variance of parameter $\beta$, we can see that the Fisher information
set the minimal possible estimate for parameter $\beta$, while higher
Fisher information means a better estimation. As the variance of $\beta$
is inverse proportional to the measurement time, we consider $N=1$
throughout this paper.

\begin{figure}
\includegraphics[width=8cm]{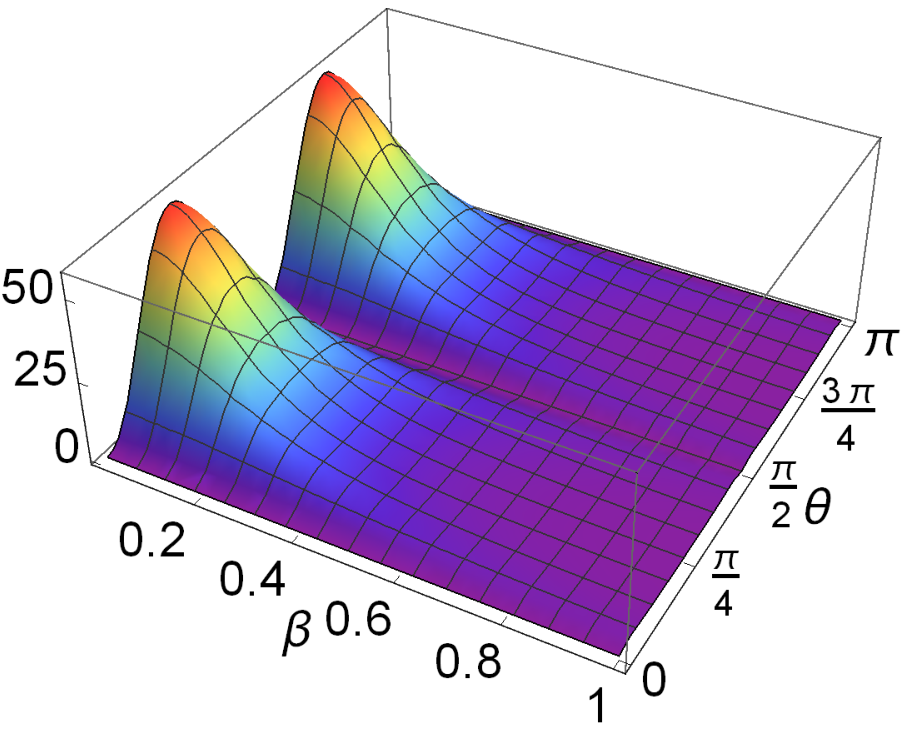}\includegraphics{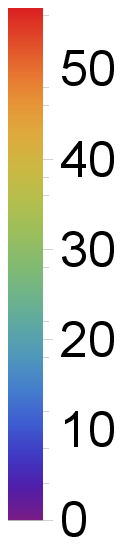}

\caption{\label{fig:Fig2}(Color online) Quantum fisher information for parameter
estimation $\beta.$ Here, we take $\phi=0,g_{0}=10^{-8},\omega_{z}=4.8\times10^{6}Hz$
, $\text{\ensuremath{\omega_{R}=3\times10^{9}Hz}}$.}
\end{figure}

\begin{figure}
\includegraphics[width=8cm]{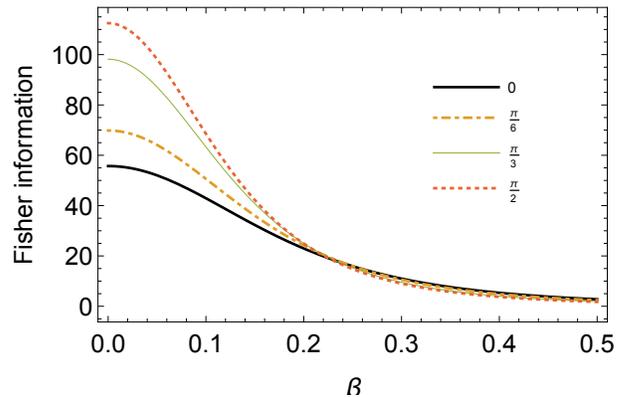}

\caption{\label{fig:Fig3}(Color online) Quantum fisher information for parameter
estimation $\beta$ with different phase of postselected states. Here
$\text{\ensuremath{\theta=\frac{\pi}{4}} }$and other parameters are
the same with Fig.\ref{fig:Fig2}. }

\end{figure}

We investigate the variation of quantum Fisher information for different
postselected states $\text{\ensuremath{\vert\psi_{f}\rangle=\cos\frac{\theta}{2}\vert\uparrow\rangle+e^{i\varphi}\sin\frac{\theta}{2}\vert\downarrow\rangle}}$
where $\theta\in[0,\pi],\varphi\in[0,2\pi)$, and $\vert\uparrow\rangle$
and $\vert\downarrow\rangle$ are eigenvector of the spin operator
$\hat{S}_{z}$ with corresponding eigenvalues $+\frac{1}{2}$ and
$-\frac{1}{2}$, respectively. Our numerical results in Fig.\ref{fig:Fig2}
and Fig.\ref{fig:Fig3} show that the quantum Fisher information is
higher in high temperature regime, and phase of postselected state
also can effect of the precision of our scheme. However, as we mentioned
in this section, when the postselected state is an eigenstate of the
spin operator $\hat{S}_{z}$, the Fisher information become zero and
our scheme lose its validity of measure the temperature of the spin.

\section{Conclusion}

In this study, we have proposed a new scheme to measure the temperature
of spin which put it in thermal bath with temperature $T$ by using
postselected weak measurement method. We find that in high temperature
regime, the precision of our proposal is high enough and can be controlled
by adjusting the parameters of postselected states of the spin. Furthermore,
in Ref.\cite{Brun}, the authors studied that quantum Fisher information
is higher in postselected rather than non-postselected weak measurement.Thus,
even though we only consider the high temperature regime, but in our
scheme if someone use the postselected measurement, it will show its
good performance in low temperature regime rather than non-postselected
measurement schemes. 
\begin{acknowledgments}
This work was supported by the National Natural Science Foundation
of China (Grant No. 11865017, No.11864042).
\end{acknowledgments}

\end{document}